# Penetration depth of Cooper pairs in the IrMn antiferromagnet

R. L. Seeger,[*,1] G. Forestier,[1] O. Gladii,[1] M. Leiviskä,[1] S. Auffret,[1] I. Joumard,[1] C. Gomez,[2] M. Rubio-Roy,[1] A. I. Buzdin,[3,4] M. Houzet,[5] and V. Baltz[**,1]

[1] Univ. Grenoble Alpes, CNRS, CEA, Grenoble INP, IRIG-Spintec, F-38000 Grenoble, France

[2] Grenoble INP, CIME Nanotech, F-38000 Grenoble, France

[3] Univ. Bordeaux, CNRS, LOMA, F-33405 Talence, France

[4] World-Class Research Center "Digital biodesign and personalized healthcare", Sechenov First Moscow State Medical University, Moscow, Russia

[5] Univ. Grenoble Alpes, CEA, Grenoble INP, IRIG, PHELIQS, F-38000 Grenoble, France

[*] rafael.lopesseeger@cea.fr

[**] vincent.baltz@cea.fr

**Abstract**

Suppression of superconductivity due to the proximity effect between a superconductor and a ferromagnet can be partially alleviated when a Cooper pair simultaneously samples different directions of the short-range exchange field. The superconductor's critical temperature, $T_C$, is therefore expected to partially recover when the ferromagnet is in a multi-domain state, as opposed to a single-domain state. Here, we discuss series of experiments performed with ferromagnet(Pt/Co)/spacer(IrMn and Pt)/superconductor(NbN) heterostructures. By tuning the various parameters in play, e.g., superconducting coherence length-to-thicknesses ratio, and domain sizes, we obtained up to 10% recovery of the superconducting critical temperature $\Delta T_C/T_C$. This large-scale recovery made novel investigations possible. In particular, from the spacer thickness-dependence of $\Delta T_C/T_C$, it was possible to deduce the characteristic length for Cooper pair penetration in an IrMn antiferromagnet. This information is crucial for electronic transport, and up to now has been difficult to access experimentally for antiferromagnets.

# I. INTRODUCTION

The interplay between superconductivity and magnetism has attracted considerable attention in recent decades [1,2] due to its importance for studies in basic physics and related applications. As a consequence, a variety of phenomena have been described in ferromagnet/superconductor hybrids, such as the spin switch effect [3–5], the superconducting magnetoresistance effect [6,7], and domain wall superconductivity [8–13]. At the heart of domain wall superconductivity, Cooper pairs consisting of electrons of opposing spins experience the short-range exchange field averaged over the superconducting coherence length. This phenomenon reduces the critical temperature ($T_C$) of the superconducting layer. A magnetic domain wall flanked by opposite spins reduces the averaged exchange field and thus allows partial recovery of the superconducting temperature, $\Delta T_C$. Recovery is achieved through the creation of an additional – and more efficient – superconducting pathway in the magnetic layer [1]. In practice, ferromagnetic domains also generate long-range dipolar magnetic fields. While nucleation of the superconductivity can occur near domain walls [8,14], dipolar fields may also cause the overall superconducting temperature to drop. This type of competition between exchange and dipolar interactions is especially significant for ferromagnets with out-of-plane anisotropy, such as [Pt/Co] multilayers [11]. Consequently, observation of the actual temperature enhancement due to the proximity effect near ferromagnetic domain walls is difficult. To overcome the inherent difficulty, the effects of two interfaces can be cumulated, for example by sandwiching a 38-nm-thick Nb superconductor between [Co(0.6)/Pt(1.5)]$_4$ and [Co(0.4)/Pt(1.1)]$_4$ (nm) ferromagnetic multilayers. This approach allowed Zhu et al. [12] to demonstrate a small ferromagnetic domain wall proximity effect of $\Delta T_C / T_C = 0.6\%$.

In antiferromagnet/superconductor heterostructures, suppression of $T_C$ [15–18] was reported with Cr and IrMn antiferromagnets, whereas Josephson current in superconductor/antiferromagnet/superconductor trilayers [19–22] was observed with an FeMn antiferromagnet. Although few studies have been published on antiferromagnet/superconductor heterostructures compared to the number available for ferromagnet/superconductor systems, they could provide crucial information on the transport properties of antiferromagnets. Indeed, these properties recently attracted interest for their use in the context of spin-dependent transport [23,24]. Understanding whether antiferromagnetic spin textures influence the transport of Cooper pairs and determining the characteristic lengths promoting transport are some of the basic points that deserve to be investigated.

In this study, we measured the proximity effect in ferromagnet(Pt/Co)/spacer(IrMn and Pt)/superconductor(NbN) heterostructures. We created domains in the ferromagnet and varied the configurations from multi- to single-domain. While controlling the domain state, we observed its influence on the superconductor's critical temperature (section II). By tuning the various parameters in play, e.g., superconducting coherence length-to-thicknesses ratio and domain sizes, we achieved recovery of the superconducting critical temperature $\Delta T_C/T_C$ by up to 10% (section III.A). This amplitude was compatible with two types of studies that were previously impossible: i) we probed the gradual evolution of $\Delta T_C/T_C$ for all the intermediate magnetic configurations of the ferromagnet (section III.B); and, ii) we demonstrated how $\Delta T_C/T_C$ decreases gradually with the thickness of the spacer layer, thus we were able to determine the penetration depth of Cooper pairs in the IrMn antiferromagnetic spacer (section III.C).

## II. EXPERIMENTS

The full stacks used in this study were (from substrate to surface): Si/SiO$_2$(500)//[Pt(1)/Co(0.65)]$_n$/spacer($t_{\text{spacer}}$)/NbN($t_{\text{NbN}}$) (nm) multilayers, where $n$ is the number of repetitions of the Pt/Co heterostructure composing the ferromagnet. The value of $n$ was varied between 4 and 25, corresponding to a variation of the nominal thickness between 6.6 and 41.25 nm. The thickness of the spacer layer (IrMn or Pt), $t_{spacer}$, was varied between 1 and 60 nm; $t_{NbN}$ is the thickness of the NbN superconducting layer and was set to 15, 30, or 60 nm. Stacks were deposited at room temperature by dc-magnetron sputtering on Si/SiO$_2$(500) (nm) substrates under argon at a pressure of 2.3 x 10$^{-3}$ mbar. The IrMn layer was deposited from an Ir$_{20}$Mn$_{80}$ (at. %) target. The superconducting NbN layers were prepared by reactive sputtering of Nb under N$_2$ gas at a partial pressure of 5 x 10$^{-3}$ mbar. The thin films were patterned into H-bars (measuring 200 µm wide and 5 mm long) by laser lithography and plasma etching. An optical image of the resulting H-bar is shown in Fig. 1(a). Electrode contacts were created using aluminum-wire bonding on 200 x 200 µm² contact pads. Electrical parameters were then measured using standard four-point current-voltage geometries, applying an ac current (lock-in detection) of amplitude 0.5 mA and frequency 13.65 Hz.

By measuring the transverse voltage between contacts $V_1$ and $V_3$ (Fig. 1(a)), the anomalous Hall contribution from the stack was determined. This contribution is known to be proportional to the perpendicular component of magnetization, $M$ [25]. Representative data showing how normalized $M$ ($m=M/M_S$) depends on an external magnetic field, $H$, applied out-of-plane for a Si/SiO$_2$//[Pt(1)/Co(0.65)]$_{15}$/IrMn(3)/NbN(15) (nm) stack are plotted in Fig. 1(b). Data-points were measured at 12 K after demagnetizing the sample by applying an alternating field of decreasing amplitude, from 10 kOe to 0 kOe at a rate of 50 Oe.s$^{-1}$. Subsequently, distinct field sequences were applied to produce different magnetic states for the Pt/Co ferromagnet. For example, the symbols in Fig. 1(b) indicate that the demagnetized (red square, for $m=M/M_S$ ~ 0) and saturated (blue circle, for $m=M/M_S$ ~ 1) states can both be accessed at a remanent field of $H$ = 0.5 kOe. These states were used throughout the study. Magnetic force microscopy (MFM) performed at room temperature (Fig. 1(d)) [26] revealed that the demagnetized state consists of maze domains, with a typical width $w_{Pt/Co}$ = 47 nm. This width was determined from the power spectral density profile of the two-dimensional Fourier transform of the MFM image (Inset in Fig. 1(d)). The domains are separated by domain walls ($\delta_{Pt/Co}$) measuring (11.5 ± 1.5) nm thick. $\delta_{Pt/Co}$ was calculated from $\delta_{Pt/Co} = \pi\sqrt{A/K}$, where A = (3.4 ± 0.4) x $10^{-7}$ erg cm$^{-2}$ and K = (1.5 ± 0.2) x $10^6$ erg cm$^{-3}$ were determined by applying the Kaplan model [27]. This model will be further discussed below. The full hysteresis loop given in Fig 1(b) is also consistent with the preferential formation of maze domains, caused by competition between exchange and magnetostatic energies [28]. It should be noted that the saturated state was only accessible here at a remanent field of 0.5 kOe. Thinner Pt/Co ferromagnets produce hysteresis loops with shape closer to a square, and can thus be used to access all states in zero applied field. Nevertheless, section III presents a discussion of some of the considerations to be taken into account when choosing thicker vs thinner Pt/Co ferromagnets for this type of study.

The superconducting critical temperature, $T_C$, of the NbN layer was determined from temperature($T$)-dependent measurements of the stacks' resistance, $R$, based on the longitudinal voltage between contacts $V_1$ and $V_2$ (Figs. 1(c) and 1(e)). To allow data comparison, $T_C$ was defined throughout as the temperature for which $R$ dropped to 0.5 mΩ, i.e., above the noise level. Typical $R$ vs $T$ measurements for H = 0.5 kOe are shown in Fig. 1(c) for [Pt/Co]$_n$/IrMn/NbN multilayers with the Pt/Co ferromagnet in a demagnetized or saturated state, and in Fig. 1(e) for a single layer of NbN subjected to the same field-cycling protocol. These data for the NbN monolayer were used to verify that the NbN superconductor is not intrinsically sensitive to field-cycling procedures. Subsequent findings could thus be confidently interpreted. Comparing Figs. 1(c) and 1(e), $T_C$ was observed to be approximately 20% smaller in the [Pt/Co]$_n$/IrMn/NbN multilayer ($T_C$ ~ 6.5 K when the Pt/Co ferromagnet is saturated) than in the monolayer of NbN ($T_{C0}$ ~ 8.4 K). This weakening of superconductivity is caused by the exchange field sampled by the Cooper pairs travelling across the spacer layer, inducing effective pair-breaking [10]. The fact that the 3-nm-thick IrMn spacer layer is transparent for the transport of Cooper pairs will be addressed specifically below. The results shown in Fig. 1(c) confirmed that the presence of domains and the resulting domain walls in the demagnetized Pt/Co ferromagnet led to weaker Cooper pair-breaking effects, as expected from the theory [10]. This effect resulted in a larger $T_C$ (~7.2 K) than that recorded for the saturated state (~ 6.5 K). Thus, relative recovery of $T_C$, defined as $\Delta T_C/T_C$ = ($T_{c,\text{demagnetized state}}$ − $T_{c,\text{saturated state}}$) / $T_{c,\text{saturated state}}$, was up to ~10% - an order of magnitude larger than the ~0.6% reported previously [12]. In fact, to observe this effect, several parameters (superconducting coherence length, $\xi_{NbN}$, vs layer thicknesses, $t_{NbN}$, $t_{IrMn}$, $t_{Pt/Co}$, vs magnetic domain width, $w_{Pt/Co}$, and domain walls width, $\delta_{Pt/Co}$) must be appropriately adjusted with respect to each other. For example: i) optimizing proximity effects requires the $\xi_{NbN}/t_{NbN}$ ratio to be maximized, but minimizing finite-size effects on superconductivity imposes a lower limit on $t_{NbN}$, or ii) optimizing the influence of a domain wall on superconductivity imposes that $\delta_{Pt/Co}$ be of the same order of magnitude as $\xi_{NbN}$. The *ad hoc* adjustment of several parameters produced the reported $\Delta T_C/T_C$, up to ~10%. Specifically, the results shown in Fig. 1 were obtained with a sample in which: $t_{NbN}$ = 15 nm, $\xi_{NbN}$ = 15 nm, $\delta_{Pt/Co}$ = 11.5 nm, $w_{Pt/Co}$, = 47 nm, $t_{Pt/Co}$ = 41.25 nm, and $t_{IrMn}$ = 3 nm. Parameter tuning will be discussed in the next section.

It is interesting to note that a similar recovery of $T_C$ was measured when using contacts $V_4$ and $V_5$ instead of $V_1$ and $V_2$ for the measurements (Fig. 1(a)), i.e., when the total number of domains probed was reduced but the overall maze arrangement remained the same. This result confirms that it is the maze arrangement that produces the observed effect.

## III. RESULTS AND DISCUSSIONS

### A. Influence of superconductor's thickness and related properties

We will now comment on the influence of the thickness of the NbN superconductor on the recovery of $T_C$ in [Pt/Co]$_n$/IrMn/NbN multilayers. The data presented in Fig. 2(a) show that $\Delta T_C$ decreases when $t_{NbN}$ increases, confirming the interfacial nature of the effect observed [10]. We gained further insights into the thickness-dependence of the NbN properties from series of measurements of $R$ vs $T$ for several applied fields. The resulting $H$-dependences of $T_C$ (Fig. 2(b)) were fitted using Ginzburg-Landau (GL) theory, which is expected to apply in the perpendicular field configuration for type II superconductors. Specifically, we deduced the superconducting coherence length, $\xi_{NbN}$, using the following equation [29]: $H = \Phi_0 \, (1 - T/T_{C,H=0})/(2\pi \xi_{GL,T=0}^2)$, where $\Phi_0$ is the magnetic flux quantum ($\Phi_0 = h/(2e)$), and $\xi_{GL}$ is the GL coherence length, with $\xi_{GL,T=0} = \xi_{NbN}\pi/2$. It should be remembered that, in the dirty limit, $\xi_{NbN} = \sqrt{\hbar D_{NbN}/(2\pi k_B T_C)}$, where $D_{NbN}$ is the diffusion constant (see also Appendix). Data derived from the fits of $H$ vs $T$ for several $t_{NbN}$ were plotted for [Pt/Co]$_n$/IrMn/NbN and NbN stacks Fig. 2(c). The corresponding $t_{NbN}$-dependences of $T_{C,H=0}$ and $\xi_{NbN}$ are known to be related to finite-size effects taking weakened interfacial superconductivity into account [29]. Most importantly, measurements of the finite-size effect on $\xi_{NbN}$ were used to produce the data presented in Fig. 2(d), where $\Delta T_C/T_C$ can be observed to scale linearly with $\xi_{NbN}/t_{NbN}$. This relationship further supports the interfacial nature of the proximity effect involved here.

### B. Influence of ferromagnetic domain configuration

As the temperature recovery observed here in [Pt/Co]$_n$/IrMn/NbN multilayers was considerable, it was possible to explore how $\Delta T_C/T_C$ evolved for several ferromagnetic configurations of the Pt/Co multilayer. Intermediate configurations, between the demagnetized and saturated states discussed in section II, were obtained as illustrated in Fig. 3(a). Specifically, an incremental sequence of minor hysteresis loops was applied at 12 K. Starting from a demagnetized state, the magnetic field was raised to $H_i$ and then reduced to 0.5 kOe. The symbols in Fig. 3(a) indicate the magnetic states that we considered. The gradual increase in $m=M/M_S$ for the intermediate states accounts for the partial remagnetization and gradual evolution of the domain configuration in the Pt/Co multilayer. After each step of the sequence in field, the $T_C$ of the superconductor was deduced from an $R$ vs $T$ scan at H = 0.5 kOe (Fig. 3(b)). The plot of $\Delta T_{C,i}/T_C$ = ($T_{c,\text{intermediate state}}$ − $T_{c,\text{saturated state}}$) / $T_{c,\text{saturated state}}$ vs $1-M/M_S$ (Fig. 3(c)) shows how the magnetic domain arrangement in the Pt/Co ferromagnet influenced superconductivity recovery in the NbN film. In particular, we observed that gradually reducing the domain size, from infinite in the saturated state ($1-M/M_S \sim 0$) to 45 nm in the demagnetized state ($1-M/M_S \sim 1$) led to progressive recovery of superconductivity, from $\Delta T_{C,i}/T_C \sim 0$ to 10%. Overall, this behavior can be explained by a theoretical model, which is detailed below.

Before coming to the model, we considered how superconductivity recovery was affected by the thickness of the Pt/Co ferromagnet. Figure 4(a) shows that the gradual increase in $\Delta T_{C,i}/T_C$ as the magnetic domain configuration of the Pt/Co multilayer shifted from saturated to demagnetized appeared to follow a universal trend that is independent of $t_{Pt/Co}$. However, the maximum value, corresponding to $\Delta T_C/T_C$ = ($T_{c,\text{demagnetized state}}$ − $T_{c,\text{saturated state}}$) / $T_{c,\text{saturated state}}$, did significantly depend on $t_{Pt/Co}$ (Fig. 4(b)), leveling out from $n$ = 15. This number of Pt/Co layers corresponds to a nominal $t_{Pt/Co}$ thickness of 24.75 nm. We note that data in Fig. 4(b) were measured for both $H$ = 0.5 kOe and 1.3 kOe, to allow exploration of larger $t_{Pt/Co}$ values. Indeed, for larger values, saturation can only be reached with a 1.3-kOe field (see section II and discussion therein). Data for $H$ = 0.5 kOe naturally show larger values than data for $H$ = 1.3 kOe as superconducting properties are weakened when a stronger field is applied. The $t_{Pt/Co}$-dependence of $\Delta T_C/T_C$ is undoubtedly driven by several parameters. First, superconducting properties are more affected by a thicker ferromagnet, as long as $t_{Pt/Co}$ remains shorter than the Cooper pair coherence length, $\xi_{Pt/Co}$. However, this effect should not be involved here, as the coherence length is known to be a few nanometers in ferromagnets. Second, since the size of the domains decreases down to a threshold thickness corresponding to about 47 nm (Fig. 4(c)), the density of domain walls increases and then levels out, resulting in a similar shape for $\Delta T_C/T_C$ vs $t_{Pt/Co}$. Specifically, the size of the domains in the demagnetized state, $w_{Pt/Co}$, changes with the number of layers, *n*, making up the [Pt/Co]$_n$ multilayer (i.e., with $t_{Pt/Co}$). The thickness-dependence of $w_{Pt/Co}$ is known to obey Kaplan's model, which accounts for the fact that the cost in domain wall energy was compensated by the gain in demagnetizing energy as the film thickness increased [27]. For $w_{Pt/Co} / t_{Pt/Co} > 1.5$, the thickness-dependence of $w_{Pt/Co}$ is given by: $Ln(w_{Pt/Co}/t_{Pt/Co}) = \pi w_0 / (2 t_{Pt/Co}) + a$, with $a = Ln(\pi) - 1 + \mu(0.5 - Ln(2))$; $\mu = 1 + 2\pi M_S^2/K$. Considering $M_S$ equal to 550 emu cm$^{-3}$ ($M_S$ = $M_{Co}t_{Co}/(t_{Co}+t_{Pt})$), data-fitting returned $w_0$ = (19.8 ± 2) nm and an anisotropy of K = (1.5 ± 0.2) x 10$^6$ erg.cm$^{-3}$. These values are in agreement with previous findings [26]. The domain wall energy $\sigma_w$ = (5.5 ± 0.5) erg.cm$^{-2}$, the exchange stiffness A = (3.4 ± 0.4) x 10$^{-7}$ erg cm$^{-2}$, and the domain wall width $\delta_{Pt/Co}$ = (11.5 ± 1.5) nm were subsequently calculated by applying the following relations: $\sigma_w = 4\sqrt{AK} = 2\pi M_S^2 w_0$ and $\delta_{Pt/Co} = \pi\sqrt{A/K}$ [27].

Because the NbN layer was grown on top of the multilayers, for which the thicknesses varied significantly, we verified that its superconducting properties were not significantly altered as a result of growth issues. $\xi_{NbN}$ and $T_{C,H=0}$ were therefore extracted using the same procedure and equations as described in Fig. 2(b) and corresponding text. These data confirmed negligible variability in NbN properties across samples (Fig. 4(c)).

We next sought to develop a theoretical model supporting the experimental findings. The model considers that Cooper pairs feel a reduced effective exchange field that is spatially uniform over the surface of the demagnetized ferromagnet. Within the quasiclassical diffusive theory for superconducting heterostructures, we derived an expression for the critical temperature of a superconductor/ferromagnet bilayer in the presence of a periodic magnetic domain structure (see Appendix). Thus, assuming a thin superconducting layer ($t_s \ll \xi_s$ with S=NbN) in good electrical contact with a thick ferromagnetic layer ($t_F \gg \xi_s$ with F=Pt/Co), and narrow domain walls ($\delta_F \ll \xi_s$), we found:

$$\frac{\Delta T_{C,i}}{\delta T_C} = \frac{T_C(m) - T_C(m=1)}{\delta T_C} =$$

$$1 - m^2 - \frac{2(L_F/\xi_s)^4}{7\pi^6 \zeta(3)} \sum_{p=1}^{\infty} \frac{\sin^2\left[\frac{\pi}{2}(m+1)p\right]}{p^6} \left[ \frac{\pi^4 p^2}{(L_F/\xi_s)^2} + \psi\left(\frac{1}{2}\right) - \psi\left(\frac{1}{2} + \frac{2\pi^2 p^2}{(L_F/\xi_s)^2}\right) \right]$$

(1)

for an arbitrary ratio $L_F/\xi_s$. Here $\zeta$ and $\psi$ are the Riemann zeta and digamma functions, respectively, and $L_F$ is the period of the domain structure ($L_F = 2w_F$). The domain structure consists of alternating majority and minority stripe domains, the relative lengths of which determine the reduced magnetization, $m=M/M_S$. The maximal shift was obtained for $L_F \ll \xi_s$, given by $\Delta T_C = \delta T_C$ with:

$$\delta T_C = \frac{7\zeta(3)}{4\pi^2} \frac{\tilde{h}^2}{k_B^2 T_{C0}}.$$

(2)

Here, $T_{C0}$ is the critical temperature of the bare superconducting layer, and $\tilde{h}$ is an effective exchange field. This field can be related to the exchange field $h$ acting on the electron spins in the ferromagnetic layer as follows:

$$\tilde{h} = \frac{\hbar \sigma_F D_S}{2\sigma_S t_S} \sqrt{\frac{h}{\hbar D_F}}$$

(3)

with the conductivities $\sigma_S$ and $\sigma_F$, and the diffusion constants $D_S$ and $D_F$ in the superconducting and ferromagnetic layers, respectively. From Fig. 4(a) it emerges that Eq. (1) for $L_F \lesssim \xi_s$, where it approximates to $\Delta T_{C,i} = (1 - m^2)\delta T_C$, qualitatively describes the experiment for $0 \leq m \ll 1$. Deviations for $1 - m \ll 1$ are attributed to the limitations of the model close to saturation, when the domain structure is very different from periodic stripes.

As the ratio $L_F/\xi_S$ increases, the $\Delta T_c$ shift is progressively reduced. Ultimately, for $L_F \gg \xi_S$, Cooper pairs mostly feel single domains, making both the demagnetized and saturated states detrimental, producing similar depairing efficiency. For $L_F \gg \xi_S$, Eq. (1) yields:

$$\Delta T_C = \frac{(2^{7/2}-1)\zeta(7/2)}{\sqrt{2}\pi^2} \frac{\tilde{h}^2}{k_B^2 T_{C0}} \frac{\xi_S}{L_F}$$

(4)

The above equation qualitatively describes the suppression of $\Delta T_C$ as the ferromagnetic layer gets thinner and $L_F = 2w_F$ concomitantly increases (see Figs. 4(b) and (c)).

As $L_F$ increases further, the assumption of a superconducting order parameter that is almost spatially uniform - used to derive Eq. (1) - breaks down. This happens when the transition takes place in the domain wall superconducting (DWS) phase, according to [10]. In this case, the increase in critical temperature between demagnetized and saturated configurations can be determined as follows:

$$\Delta T_C = \frac{(8\sqrt{2}-1)^2 \zeta^2(7/2)}{8\pi^6} \frac{\tilde{h}^4}{k_B^4 T_{C0}^3}$$

(5)

This relationship requires the walls to be sufficiently distant from each other on the characteristic DWS length scale [10], i.e., $(1-m)L_F \geq \xi_{DWS}(T)$, with $\xi_{DWS}(T) = \xi_S T_{C0}/(T_{C0}-T)$. As $T_{C0} - T \approx \tilde{h}^2/(k_B^2 T_{C0})$ in this regime, Eqs. (4) and (5) can be seen to match parametrically at $m = 0$ (i.e., $L_F \approx \xi_{DWS}(T_C)$). Note that the DWS phases overlap extensively at $L_F \approx \xi_{GL}(T_C)$, where $\xi_{GL}(T) = \frac{\pi}{2}\xi_S\sqrt{T_{C0}/(T_{C0}-T)}$, is the GL coherence length.

### C. Influence of the nature, thickness, and domain state of the spacer layer

We next investigated the influence of the spacer thickness on the recovery of superconductivity. More particularly, we took advantage of the proximity effect in our ferromagnet/antiferromagnetic-spacer/superconductor heterostructures to study the transport properties of Cooper pairs in the IrMn antiferromagnet and to further deduce characteristic properties that could be of interest for any electronic transport-related study, e.g., for antiferromagnetic spintronics [23,24]. How $\Delta T_C / T_C$ depends on the IrMn spacer thickness is shown in Fig. 5(a) for two values of applied field. It should be noted that the superconducting properties of NbN, $\xi_{NbN}$ and $T_{C,H=0}$ were tested in this set of samples. These data indicated that the variability in $\xi_{NbN}$ and $T_{C,H=0}$ across samples was negligible (not shown), in line with the data presented in Fig. 4(c) and the related discussion. The overall reduction of $\Delta T_C/T_C$ with $t_{IrMn}$ (Fig. 5(a)) relates to the coherence length of Cooper pairs in the metallic spacer of the IrMn antiferromagnet, $\xi_{IrMn}$. The fact that the overall signal only entirely vanished when the $t_{IrMn}$ thickness reached ~ 40 nm indicates that a thin IrMn layer (e.g., 3 nm as considered in the previous sections) will be completely transparent for the electronic transport of Cooper pairs. This finding can be explained by the fact that an antiferromagnetic exchange length of a few nanometers is much shorter than the superconducting coherence length of a few tens of nanometers. As a result, the different directions of the moments are sampled simultaneously by a Cooper pair, and the antiferromagnet is viewed as a non-magnetic layer in the Cooper pair reference frame. We note that the spin structure in polycrystalline IrMn thin films like the ones used in our samples resembles a disordered phase ($\gamma$-phase) of the non-collinear structure of the bulk $L1_2$-IrMn$_3$ antiferromagnet [30]. When considering the diffusion of Cooper pairs, we took $\Delta T_C/T_C \propto \exp[-k_{spacer}t_{spacer}]$ - expected from quasiclassical theories in the diffusive limit - with a wavevector of the form $k_{spacer} = 1/\xi_{spacer}$. Fitting these relations to the data shown in Fig. 5, we obtained a coherence length of $\xi_{IrMn} = (6.7 \pm 1)$ nm for the IrMn antiferromagnet. In comparison, a value of $(12.4 \pm 2)$ nm was obtained for the non-magnetic Pt layer. The expected result was $\xi_{spacer} \propto \sqrt{D_{spacer}\tau_{spacer}}$, where $D_{spacer}$ is the electron diffusion constant, and $\tau_{spacer}$ is the depairing time for Cooper pairs in the metallic spacer layer, which includes contributions from spin relaxation processes [31].

We finally considered whether the magnetic state of the IrMn antiferromagnet influenced superconductivity. We took advantage of the strong exchange bias interaction between the IrMn bottom interface and the adjacent Pt/Co ferromagnet to imprint ferromagnetic configurations in the IrMn antiferromagnet [32]. Initially, exchange bias interaction was quenched by raising the sample's temperature to 400 K, i.e., above the blocking temperature ($T_B$) for the ferromagnet/antiferromagnet bilayer (Fig. 6(a)). In these conditions, the IrMn antiferromagnetic layer lost its ability to pin the magnetization of the adjacent Pt/Co ferromagnet. Consequently, this layer can be considered to be a single-layer ferromagnet in which different magnetic state types – demagnetized, saturated, or any intermediate state – can be nucleated by conventional means (see Fig. 3(a) and corresponding text). Subsequently, the bilayer was cooled below $T_B$, (here, down to T = 12 K), causing the moments in the antiferromagnet to align with those of the ferromagnet due to exchange bias coupling. Indeed, below the blocking temperature, the moments in the antiferromagnet remained pinned regardless of the direction of the moments in the ferromagnet; at 12 K this effect produced a hysteresis loop shift, $H_E$. This procedure was demonstrated to be robust and has been used elsewhere to imprint multi-domain states and magnetic textures in antiferromagnets. Thus, for example, exchange bias was shown to allow several spin arrangements to be imprinted across the core of antiferromagnets, at least across 8 nm for exchange springs in IrMn layers [33], or 3 nm for textures such as vortices in CoO and NiO layers [34]. Figure 6(a) shows the blocking temperature distribution for a Si/SiO$_2$//[Pt(1)/Co(0.65)]$_4$/IrMn(3)/NbN(15) (nm) stack. This distribution was obtained following a proven specific process according to which $H_E$ is recorded after each step in an incremental field-cooling procedure starting from an annealing temperature $T_a$ (inset). The procedure is extensively described elsewhere [35,36]. Most importantly, the data presented in Fig. 6(a) indicate that the magnetic configuration of the Pt/Co ferromagnet can be stabilized in the IrMn antiferromagnet by cooling from 400 K down to 12 K, as the whole distribution of blocking temperatures was measured below 400 K. Using the domain replication approach mentioned above, we stabilized several states at T = 12 K in the IrMn antiferromagnet of a Si/SiO$_2$//[Pt(1)/Co(0.65)]$_{15}$/IrMn(3)/NbN(15) (nm) stack (Fig. 6(b)). This stabilization made it possible to obtain a hysteresis loop for which the shift along the *H*-axis depended on the magnetic state of the antiferromagnet [32]. For every antiferromagnetic state (pinned at 12 K), we reproduced the procedure detailed in Fig. 3(a) and related text, using sequences of minor hysteresis loops, driven by H$_i$, to scan $\Delta T_C/T_C$ for several ferromagnetic configurations, from demagnetized to saturated. Figure 6(c) shows the corresponding normalized magnetization ($m=M/M_S$) vs H$_i$. The shift observed on these curves is a direct consequence of the fact that the IrMn antiferromagnet was prepared in three distinct states. Figure 6(d) shows the gradual enhancement of $\Delta T_C/T_C$ as a function of the magnetic domain configuration of the Pt/Co for several IrMn arrangements. These results demonstrate that the recovery of superconductivity, driven by the ferromagnetic configuration of the Pt/Co multilayer (Fig. 3 and corresponding text), is independent of the domain arrangement in the IrMn antiferromagnet. This finding is consistent with the fact that a 3-nm-thick IrMn layer is transparent for the electronic transport of Cooper pairs, due to simultaneous sampling of the different directions of the moments.

## IV. CONCLUSION

In conclusion, the main contribution of this paper is that it presents a systematic investigation of the superconducting proximity effect in ferromagnet(Pt/Co)/spacer(IrMn and Pt)/superconductor(NbN) heterostructures. The findings presented indicate that by tuning the various parameters in play, the recovery of the superconducting critical temperature in the presence of ferromagnetic domains and domain walls can be maximized to a degree that makes it possible to carry out two types of studies that were previously impossible. We were therefore able to: i) probe how the recovery of the superconducting critical temperature gradually evolves with all the intermediate magnetic configurations of the ferromagnet; and, ii) determine that the recovery of the superconducting critical temperature gradually reduces with the thickness of the metallic spacer layer. Most importantly, these experiments allowed us to evaluate the penetration of Cooper pairs in the IrMn metallic antiferromagnet, information which is crucial for electronic transport, and up to now has been difficult to access experimentally for antiferromagnets. The results presented therefore open a new pathway for the investigation of electronic transport in antiferromagnetic materials for spintronics.

**ACKNOWLEDGMENTS**

The authors acknowledge financial support from the French national research agency (ANR) [Grant Number ANR-15-CE24-0015-01], KAUST [Grant Number OSR-2015-CRG4-2626], and the CEA's bottom-up exploratory program (Grant Number PE-18P31-ELSA). A.I.B. acknowledges support by the Ministry of Science and Higher Education of the Russian Federation within the framework of state funding for the creation and development of World-Class Research Center "Digital biodesign and personalized healthcare" N075-15-2020-92. We also thank M. Gallagher-Gambarelli for critical reading of the manuscript.

**APPENDIX**

In this Appendix we provide the derivation of Eq. (1).

**A. Formalism**

Within the quasiclassical diffusive approximation, the critical temperature $T_C$ of the superconducting(S)/normal(N)/ferromagnetic(F) trilayer is obtained as the solution of the following self-consistency equation:

$$\Delta(\mathbf{r}) = \pi T \lambda(\mathbf{r}) \sum_\omega f_\omega(\mathbf{r})$$

(6)

Here $\Delta(\mathbf{r})$ is the superconducting order parameter at position $\mathbf{r} = (x, y, z)$, $\omega = (2p + 1)\pi T$ ($p$ integer) are Matsubara frequencies at temperature $T$, and $\lambda(\mathbf{r})$ is the pairing constant that takes value $\lambda$ in the S layer, and vanishes otherwise. (We use units with $\hbar = k_B = 1$ throughout the Appendix). In the S layer ($-t_S < y < 0$), the anomalous component of the quasiclassical Green function, $f_\omega(\mathbf{r})$, solves the Usadel equation:

$$-\frac{D_S}{2}\nabla^2 f_\omega(\mathbf{r}) + |\omega| f_\omega(\mathbf{r}) = \Delta(\mathbf{r})$$

(7)

where $t_S$ is the thickness of the S layer and $D_S$ is its diffusion constant.

In the N layer ($0 < y < t_N$), $f_\omega(\mathbf{r})$ solves:

$$-\frac{D_N}{2}\nabla^2 f_\omega(\mathbf{r}) + \left(|\omega| + \frac{1}{\tau_N}\right) f_\omega(\mathbf{r}) = 0$$

(8)

where $t_N$ is the thickness of the N layer, $D_N$ is its diffusion constant and $\frac{1}{\tau_N}$ is a depairing rate.

In the F layer ($t_N < y < t_N + t_F$), $f_\omega(\mathbf{r})$ solves:

$$-\frac{D_F}{2}\nabla^2 f_\omega(\mathbf{r}) + \left(|\omega| + \frac{1}{\tau_F} + ih(\mathbf{r})sign(\omega)\right) f_\omega(\mathbf{r}) = 0$$

(9)

where $t_F$ is the thickness of the F layer, $D_F$ is its diffusion constant, $\frac{1}{\tau_F}$ is another depairing rate, and $h(\mathbf{r})$ is an exchange field. Equations (7)-(9) are supplemented with boundary conditions, which express the absence of current at the interfaces with vacuum:

$$\partial_y f_\omega(x, -t_S, z) = 0$$

and

$$\partial_y f_\omega(x, t_N + t_F, z) = 0$$

(10)

as well as current conservation at each interface between two layers:

$$f_\omega(x, 0^-, z) = f_\omega(x, 0^+, z)$$

$$\sigma_S \, \partial_y f_\omega(x, 0^-, z) = \sigma_N \, \partial_y f_\omega(x, 0^+, z)$$

and

$$f_\omega(x, t_N^-, z) = f_\omega(x, t_N^+, z)$$

$$\sigma_N \, \partial_y f_\omega(x, t_N^-, z) = \sigma_F \, \partial_y f_\omega(x, t_N^+, z)$$

(11)

Here $\sigma_S$, $\sigma_N$, and $\sigma_F$ are the conductivities in S, N, and F layers, respectively, and we assumed a negligible tunnel resistance between two layers.

The exchange field in the F layer is generated by a periodic structure of magnetic domains with period $L_F$ along the $x$-direction. Assuming narrow domain walls, we approximate the exchange field as:

$$h(\mathbf{r}) = h(x) = \begin{cases} h & \text{for } 0 < x < a \\ -h & \text{for } a < x < L_F \end{cases}$$

(12)

Here $h$ is the exchange field inside a domain with (saturation) magnetization $M_S$, and the relative length of majority and minority domains is related with the reduced magnetization, $m = M/M_S = 2a/L_F - 1$, where $M$ is the magnetization. For this magnetic texture, the $z$-dependence drops out from all above equations.

To proceed further, we assume that the S layer is thin on the scale of the superconducting coherence length, $\xi_S = \sqrt{D_S/(2\pi T_{C0})}$, where $T_{C0}$ is the bare critical temperature of the S material. Then $f$ hardly varies along the $y$-direction in that layer. We find that Eq. (7) averaged over the thickness of the S layer yields:

$$-\frac{D_S}{2} \partial_x^2 f_\omega(x) - \frac{D_S}{2 t_S} \partial_y f_\omega(x) + |\omega| f_\omega(x) = \Delta(x)$$

(13)

where we used integration by parts and boundary condition of Eq. (10). The relation between $\partial_y f_\omega(x)$ and $f_\omega(x)$ is obtained by solving the Usadel equations in N and F layers.

For this, we also assume that the F layer is thick on the scale of the ferromagnetic coherence $\xi_F = \sqrt{D_F/h}$, and $\frac{1}{\tau_F}, h \gg T$ such that the term proportional to $|\omega|$ in Eq. (9) can be dropped out. Then, the solution of Eq. (9) approximately satisfies:

$$\frac{\partial_y f(x, t_N^+)}{f(x, t_N^+)} = -\left(\frac{1}{\xi_{F1}} + \frac{i}{\xi_{F2}} \frac{h(x)}{h} sign(\omega)\right)$$

(14)

with

$$\xi_{F1,F2} = \xi_F \sqrt{\frac{\sqrt{(h\tau_F)^2 + 1} \mp 1}{h\tau_F}}$$

(15)

In particular, $\xi_{F1,F2} \approx \xi_F$ if $h\tau_F \gg 1$.

We are left with solving Eq. (8) in the N layer. In the following, we consider the case of a thin N layer.

### B. Thin N layer

By using the boundary conditions Eqs. (11), and (14), for a thin N layer, Eq. (13) simplifies into:

$$-\frac{D_S}{2}\partial_x^2 f_\omega(x) + \left(|\omega| + \frac{1}{\tilde{\tau}} + i\tilde{h}(x)sign(\omega)\right)f_\omega(x) = \Delta(x)$$

(16)

with

$$\tilde{h}(x) = \tilde{h}\frac{h(x)}{h}, \quad \tilde{h} = \frac{D_S \sigma_F}{2 t_S \sigma_S \xi_{F2}} = \frac{1}{\tilde{\tau}}\frac{\xi_{F1}}{\xi_{F2}}$$

(17)

The critical temperature is then obtained from the linearized self-consistency equation:

$$\Delta(x)\ln\frac{T_{C0}}{T} = \pi T \sum_\omega \left[\frac{\Delta(x)}{|\omega|} - f_\omega(x)\right]$$

(18)

where we used a standard regularization procedure to trade λ with $T_{C0}$ in Eq. (6).

We consider the regime $\tilde{h}, 1/\tilde{\tau} \ll T_{C0}$ for which the critical temperature is weakly suppressed, $T_{C0} - T_C \ll T_{C0}$. We also consider the case of a dense domain structure, such that the order parameter at the transition into the superconducting state is almost uniform. This approximation is discussed in section III.B. By expressing Eq. (16) in Fourier space, and treating perturbatively the Fourier components of $f$ and $\Delta$ with q ≠ 0 (similar to Ref. [37]), we find:

$$f_0 \approx \frac{\Delta_0}{|\omega| + \frac{1}{\tilde{\tau}} + i\tilde{h}_0 + \sum_{q \neq 0}|\tilde{h}_q|^2/[D_S q^2/2 + |\omega|]}$$

(19)

Here $\tilde{h}_0 = m\tilde{h}$ and $|\tilde{h}_q|^2 = [4\tilde{h}\sin(qa/2)/qL_F]^2$ with $q = 2\pi p/L_F$ for the domain texture described by Eq. (12). The self-consistency Eq. (18) yields the shift of critical temperature:

$$\frac{T_{C0} - T_C}{T_{C0}} = 2\pi T_{C0} \sum_{\omega>0} \left[\frac{1}{\tilde{\tau}\omega^2} + \frac{\tilde{h}_0^2}{\omega^3} + \sum_{q \neq 0} \frac{|\tilde{h}_q|^2}{\omega^2[\omega + D_S q^2/2]}\right]$$

(20)

Equation (20) is alternatively expressed as Eq. (1) in the main text, while Eq. (17) simplifies to Eq. (3) for $h\tau_F \gg 1$.

**REFERENCES**

[1] A. I. Buzdin, *Proximity Effects in Superconductor-Ferromagnet Heterostructures*, Rev. Mod. Phys. **77**, 935 (2005).

[2] J. Linder and J. W. A. Robinson, *Superconducting Spintronics*, Nat. Phys. **11**, 307 (2015).

[3] L. R. Tagirov, *Low-Field Superconducting Spin Switch Based on a Superconductor/Ferromagnet Multilayer*, Phys. Rev. Lett. **83**, 2058 (1999).

[4] J. Y. Gu, J. S. Jiang, J. Pearson, and S. D. Bader, *Magnetization-Orientation Dependence of the Superconducting Transition Temperature in the Ferromagnet-Superconductor-Ferromagnet System : CuNi = Nb = CuNi*, Phys. Rev. Lett. **89**, 267001 (2002).

[5] A. Potenza and C. H. Marrows, *Superconductor-Ferromagnet CuNi/Nb/CuNi Trilayers as Superconducting Spin-Valve Core Structures*, Phys. Rev. B **71**, 180503 (2005).

[6] R. Steiner and P. Ziemann, *Magnetic Switching of the Superconducting Transition Temperature in Layered Ferromagnetic/Superconducting Hybrids: Spin Switch versus Stray Field Effects*, Phys. Rev. B **74**, 094504 (2006).

[7] D. Stamopoulos, E. Manios, and M. Pissas, *Synergy of Exchange Bias with Superconductivity in Ferromagnetic- Superconducting Layered Hybrids: The Influence of in-Plane and out-of-Plane Magnetic Order on Superconductivity*, Supercond. Sci. Technol. **20**, 1205 (2007).

[8] Y. Aladyshkin, A. Fraerman, S. Mel'nikov, A. Ryzhov, V. Sokolov, and I. Buzdin, *Domain-Wall Superconductivity in Hybrid Superconductor-Ferromagnet Structures*, Phys. Rev. B **68**, 184508 (2003).

[9] A. Y. Rusanov, M. Hesselberth, and J. Aarts, *Enhancement of the Superconducting Transition Temperature in Nb = Permalloy Bilayers by Controlling the Domain State of the Ferromagnet*, Phys. Rev. Lett. **93**, 057002 (2004).

[10] M. Houzet and A. I. Buzdin, *Theory of Domain-Wall Superconductivity in Superconductor/Ferromagnet Bilayers*, Phys. Rev. B **74**, 214507 (2006).

[11] A. Singh, C. Sürgers, and H. Löhneysen, *Superconducting Spin Switch with Perpendicular Magnetic Anisotropy*, Phys. Rev. B **75**, 024513 (2007).

[12] L. Y. Zhu, T. Y. Chen, and C. L. Chien, *Altering the Superconductor Transition Temperature by Domain-Wall Arrangements in Hybrid Ferromagnet-Superconductor Structures*, Phys. Rev. Lett. **101**, 017004 (2008).

[13] M. Z. Cieplak, Z. Adamus, M. Kończykowski, L. Y. Zhu, X. M. Cheng, and C. L. Chien, *Tuning Vortex Confinement by Magnetic Domains in a Superconductor/ Ferromagnet Bilayer*, Phys. Rev. B **87**, 014519 (2013).

[14] Z. Yang, M. Lange, A. Volodin, R. Szymczak, and V. V. Moshchalkov, *Domain-Wall Superconductivity in Superconductor-Ferromagnet Hybrids*, Nat. Mater. **3**, 793 (2004).

[15] Y. Cheng and M. B. Stearns, *Superconductivity of Nb / Cr Multilayers*, J. App. Phys. **67**, 118 (1990).

[16] M. Hübener, D. Tikhonov, I. AGarifullin, K. Westerholt, and H. Zabel, *The Antiferromagnet / Superconductor Proximity Effect in Cr / V / Cr Trilayers M H Ubener*, J. Phys. Condens. Matter **14**, 8687 (2002).

[17] B. L.Wu, Y. M. Yang, Z. B. Guo, Y. H.Wu, and J. J. Qiu, *Suppression of*

*Superconductivity in Nb by IrMn in IrMn / Nb Bilayers*, Appl. Phys. Lett. **109**, 152602 (2013).

[18] C. Bell, J. Tarte, G. Burnell, W. Leung, D. J. Kang, and G. Blamire, *Proximity and Josephson Effects in Superconductor/Antiferromagnetic Nb/γ-Fe50Mn50 Heterostructures*, Phys. Rev. B **68**, 144517 (2003).

[19] I. V. Bobkova, P. J. Hirschfeld, and Y. S. Barash, *Spin-Dependent Quasiparticle Reflection and Bound States at Interfaces with Itinerant Antiferromagnets*, Phys. Rev. Lett. **94**, 037005 (2005).

[20] B. M. Andersen, I. V. Bobkova, P. J. Hirschfeld, and Y. S. Barash, *0-π Transitions in Josephson Junctions with Antiferromagnetic Interlayers*, Phys. Rev. Lett. **96**, 117005 (2006).

[21] P. Komissinskiy, G. A. Ovsyannikov, I. V. Borisenko, Y. V. Kislinskii, K. Y. Constantinian, A. V. Zaitsev, and D. Winkler, *Josephson Effect in Hybrid Oxide Heterostructures with an Antiferromagnetic Layer*, Phys. Rev. Lett. **99**, 017004 (2007).

[22] C. Bell, J. Tarte, G. Burnell, W. Leung, D. J. Kang, and G. Blamire, *Proximity and Josephson Effects in Superconductor/Antiferromagnetic Nb/γ-Fe50Mn50heterostructures*, Phys. Rev. B **68**, 144517 (2003).

[23] V. Baltz, A. Manchon, M. Tsoi, T. Moriyama, T. Ono, and Y. Tserkovnyak, *Antiferromagnetic Spintronics*, Rev. Mod. Phys. **90**, 015005 (2018).

[24] T. Jungwirth, X. Marti, P. Wadley, and J. Wunderlich, *Antiferromagnetic Spintronics*, Nat. Nanotechnol. **11**, 231 (2016).

[25] C. Canedy, X. Li, and G. Xiao, *Large Magnetic Moment Enhancement and Extraordinary Hall Effect in Co/Pt Superlattices*, Phys. Rev. B **62**, 508 (2000).

[26] V. Baltz, A. Marty, B. Rodmacq, and B. Dieny, *Magnetic Domain Replication in Interacting Bilayers with Out-of-Plane Anisotropy: Application to Co/Pt Multilayers*, Phys. Rev. B **75**, 014406 (2007).

[27] B. Kaplan and G. A. Gehring, *The Domain Structure in Ultrathin Magnetic Films*, J. Magn. Magn. Mater. **128**, 111 (1993).

[28] P. D. Johnson, *Reports on Progress in Physics Related Content*, Rep. Prog. Phys. **60**, 1217 (1997).

[29] P. G. de Gennes, *Superconductivity of Metals and Alloys* (Cambridge, MA : Perseus, 1999).

[30] A. Kohn, A. Kovács, R. Fan, G. J. McIntyre, R. C. C. Ward, and J. P. Goff, *The Antiferromagnetic Structures of IrMn3 and Their Influence on Exchange-Bias.*, Sci. Rep. **3**, 2412 (2013).

[31] F. S. Bergeret, M. Silaev, P. Virtanen, and T. T. Heikkilä, *Colloquium: Nonequilibrium Effects in Superconductors with a Spin-Splitting Field*, Rev. Mod. Phys. **90**, 41001 (2018).

[32] S. Brück, J. Sort, V. Baltz, S. Suriñach, J. S. Muñoz, B. Dieny, M. D. Baró, and J. Nogués, *Exploiting Length Scales of Exchange-Bias Systems to Fully Tailor Double-Shifted Hysteresis Loops*, Adv. Mater. **17**, 2978 (2005).

[33] G. Salazar-Alvarez, J. J. Kavich, J. Sort, A. Mugarza, S. Stepanow, A. Potenza, H. Marchetto, S. S. Dhesi, V. Baltz, B. Dieny, A. Weber, L. J. Heyderman, J. Nogues, and P. Gambardella, *Direct Evidence of Imprinted Vortex States in the Antiferromagnet of*

*Exchange Biased Microdisks*, Appl. Phys. Lett. **95**, 012510 (2009).

[34] J. Wu, D. Carlton, J. S. Park, Y. Meng, E. Arenholz, A. Doran, A. T. Young, A. Scholl, C. Hwang, H. W. Zhao, J. Bokor, and Z. Q. Qiu, *Direct Observation of Imprinted Antiferromagnetic Vortex States in CoO/Fe/Ag(001) Discs*, Nat. Phys. **7**, 303 (2011).

[35] S. Soeya, T. Imagawa, K. Mitsuoka, and S. Narishige, *Distribution of Blocking Temperature in Bilayered Ni81Fe19/Ni0 Films*, J. Appl. Phys. **76**, 5356 (1994).

[36] V. Baltz, B. Rodmacq, A. Zarefy, L. Lechevallier, and B. Dieny, *Bimodal Distribution of Blocking Temperature in Exchange-Biased Ferromagnetic/Antiferromagnetic Bilayers*, Phys. Rev. B **81**, 052404 (2010).

[37] A. I. Buzdin and Bulaevskii L. N., *Ferromagnetic Film on the Surface of a Superconductor: Possible Onset of Inhomogeneous Magnetic Ordering*, Sov. Phys. JETP **67**, 576 (1988).

**FIGURE CAPTIONS**

Fig. 1. (color online) (a) Optical image of a typical device used to perform transport measurements. The complete image is reconstructed from three optical images (indicated by the dotted squares). (b) Normalized magnetization $M/M_S$ starting from a demagnetized state measured at 12 K for a Si/SiO$_2$//[Pt(1)/Co(0.65)]$_{15}$/IrMn(3)/NbN(15) (nm) ferromagnetic/spacer/superconductor stack. (c) Representative data showing the $T$-dependence of $R$, for the same sample as in (b), prepared in two distinct magnetic states: saturated and demagnetized, through two procedures involving field-cycling and cooling (see text). $\Delta T_C$ (here, systematically measured at $R = 0.5$ m$\Omega$) represents the difference in superconducting critical temperature between the saturated and demagnetized states ($\Delta T_C = T_{c,\text{demagnetized state}} - T_{c,\text{saturated state}}$). (d) MFM image taken at room temperature, for the sample used in (b) and (c), showing maze domains after demagnetization. (Inset) PSD profile of the MFM image. (e) Control experiment with a bare Si/SiO$_2$//NbN(15) (nm) stack subjected to the two procedures used for (c). Data in (c) and (e) were measured for an applied field $H = 0.5$ kOe. The symbols in (b) represent the two magnetic states, demagnetized (square) and saturated (circle).

Fig. 2. (color online) (a) NbN thickness ($t_{\text{NbN}}$) dependence of $\Delta T_C$. (b) Representative data showing $H$ vs. $T_C$ as measured for Si/SiO$_2$//[Pt(1)/Co(0.65)]$_{15}$/IrMn(3)/NbN($t_{\text{NbN}}$) (nm) stacks. These data were used to calculate the superconducting coherence length, $\xi_{\text{NbN}}$ and the zero-field superconducting temperature, $T_{C,H=0}$. Lines were fitted to the data using the model described in the text. (c) Corresponding $t_{\text{NbN}}$-dependences of $\xi_{\text{NbN}}$ and $T_{C,H=0}$ compared to data obtained for bare Si/SiO$_2$//NbN($t_{\text{NbN}}$) (nm) stacks. The lines serve as visual guides. (d) Dependence of $\Delta T_C/T_C = (T_{c,\text{demagnetized state}} - T_{c,\text{saturated state}}) / T_{c,\text{saturated state}}$ with the superconducting coherence

length to thickness ratio ($\xi_{NbN}/t_{NbN}$). The line is a linear fit to the data constrained to pass through (0,0).

Fig. 3. (color online) (a) Normalized magnetization $M/M_S$ of a Si/SiO$_2$//[Pt(1)/Co(0.65)]$_{15}$/IrMn(3)/NbN(15) (nm) stack for field cycling series. Measurements were performed at 12 K starting from a demagnetized state. Selected magnetic configurations are labeled as follows: (square) Saturated; (triangle) Intermediate; and (diamond) Demagnetized. The cooling procedures used to access each magnetic configuration are described in the text. (b) Data showing the $T$-dependences of $R$ for the three magnetic states examined. (c) Dependence of $\Delta T_{C,i}/T_C$ on 1-$M/M_S$, where $\Delta T_{C,i}$ represents the difference in superconducting critical temperature between any state and the saturated state.

Fig. 4. (color online) (a) Dependence of $\Delta T_{C,i}/\Delta T_C$ on 1-$M/M_S$ for Si/SiO$_2$//[Pt(1)/Co(0.65)]$_n$/IrMn(3)/NbN(15) (nm) stacks ($n$ = 4, 8, 11, 15), measured while applying an external field of H = 0.5 kOe. Lines were calculated using the model described in the text, for $L_F/\xi_S \approx$ 0.25; 2.5; 5; 12.5; and 25. (b) Corresponding dependence of $\Delta T_C/T_C$, corresponding to (T$_{c,\text{demagnetized state}}$ − T$_{c,\text{saturated state}}$) / T$_{c,\text{saturated state}}$) on the total thickness (t$_{Pt/Co}$) of the [Pt/Co]$_n$ multilayer, measured at H = 0.5 and 1.3 kOe. (c) t$_{Pt/Co}$-dependence of the domain sizes ($w_{Pt/Co}$), deduced from MFM images taken at room temperature after demagnetization, ie for $M/M_S$ ~ 0. (Inset) Semilogarithmic-scale dependence of $w_{Pt/Co}$ on 1/t$_{Pt/Co}$. Lines were fitted to the data using a model described in the text. (d) Control measurements for $\xi_{NbN}$ and $T_{C,H=0}$ vs t$_{Pt/Co}$.

Fig. 5. (color online) (a-b) $t_{spacer}$-dependence of $\Delta T_C/T_C$ measured for Si/SiO$_2$//[Pt(1)/Co(0.65)]$_{15}$/spacer($t_{spacer}$)/NbN(15) (nm) stacks, for IrMn and Pt spacers. Lines

correspond to exponential fits of the data (see text). (Inset) Corresponding semilogarithmic scale dependence of $\Delta T_C/T_C$ on $t_{spacer}$.

Fig. 6. (color online) (a) Blocking temperature distribution measured for a Si/SiO$_2$//[Pt(1)/Co(0.65)]$_4$/IrMn(3)/NbN(15) (nm) stack. (b) Normalized magnetization $M/M_S$, measured at 12 K for a Si/SiO$_2$//[Pt(1)/Co(0.65)]$_{15}$/IrMn(3)/NbN(15) (nm) stack, after stabilizing several states in the IrMn antiferromagnet (see text). (c) Corresponding normalized magnetization at the remanent state for H = 0.5 kOe. (d) Dependence of $\Delta T_{C,i}/T_C$ on $1-M/M_S$.

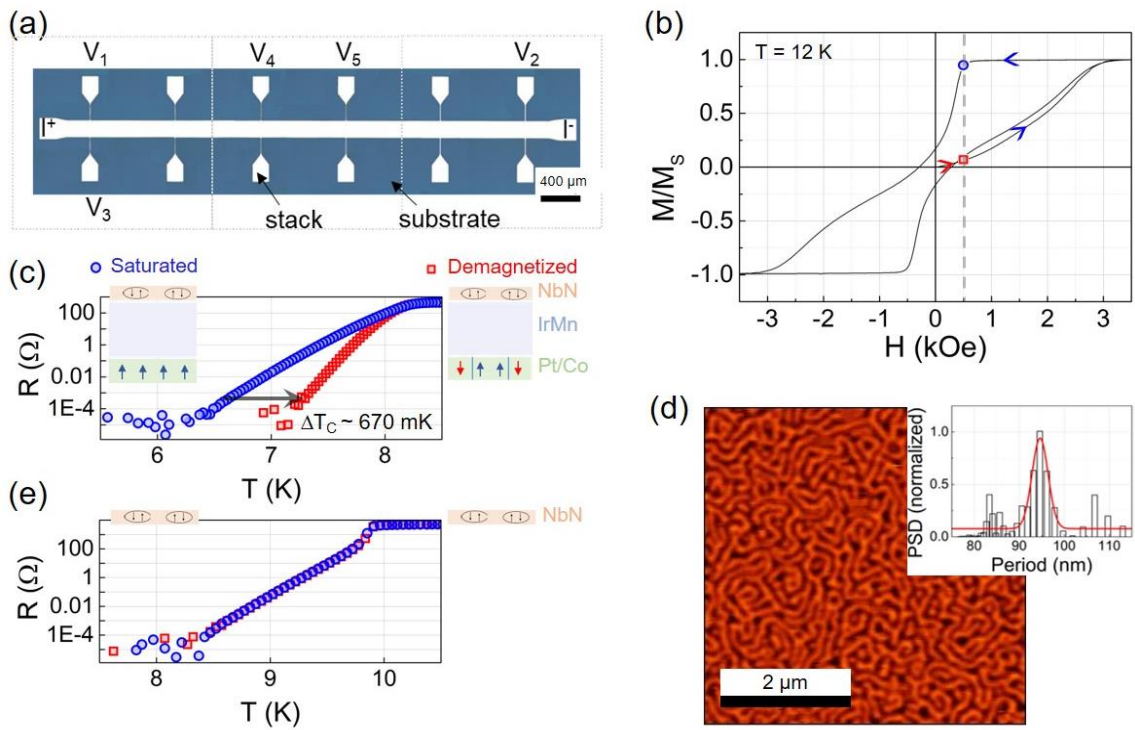

Fig. 1

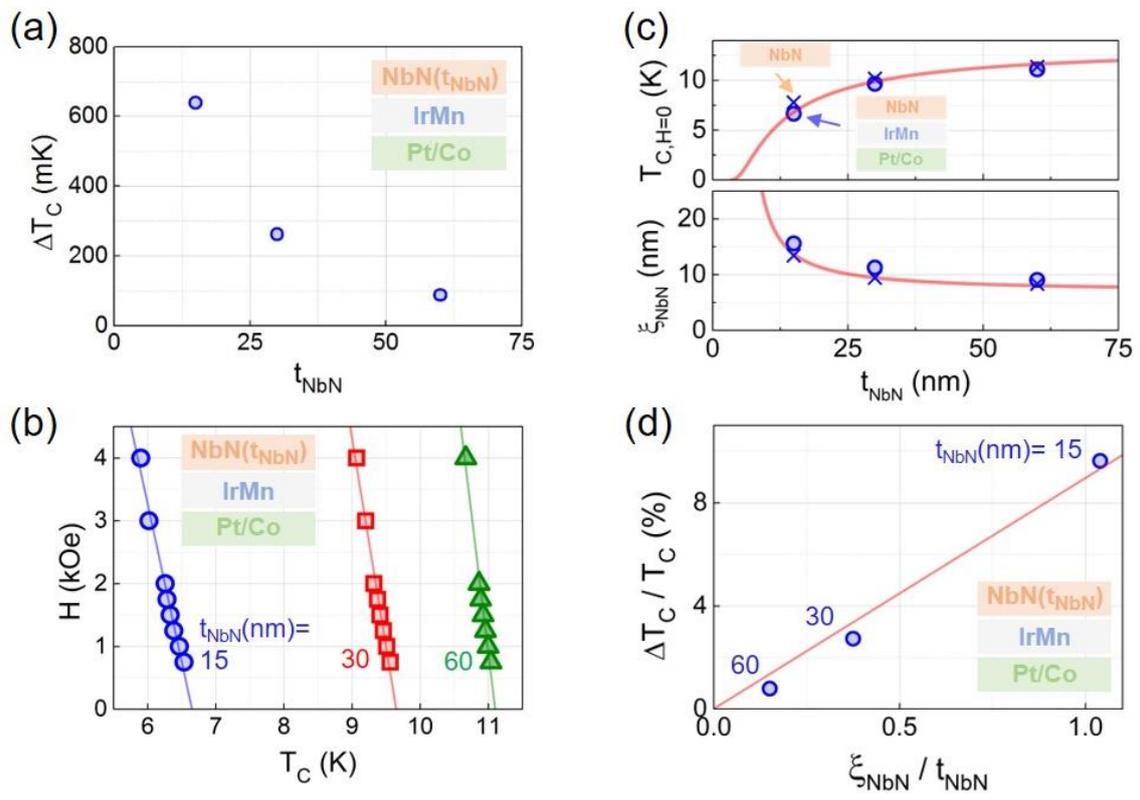

Fig. 2

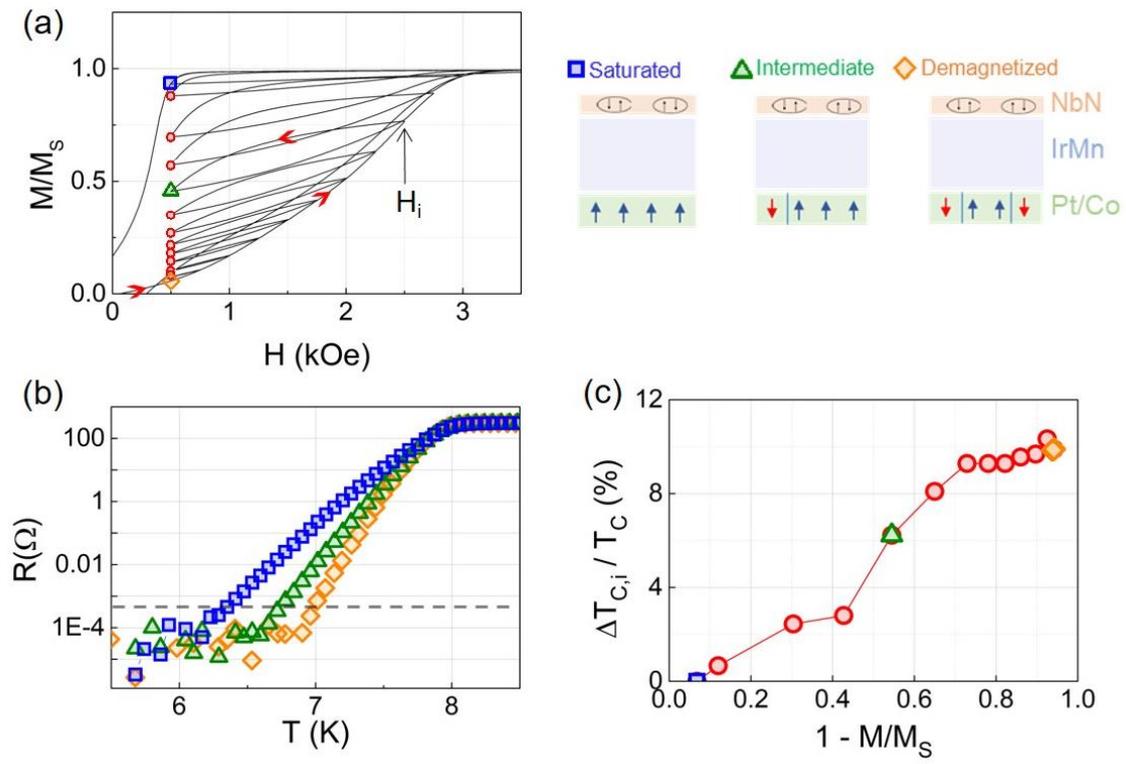

Fig. 3

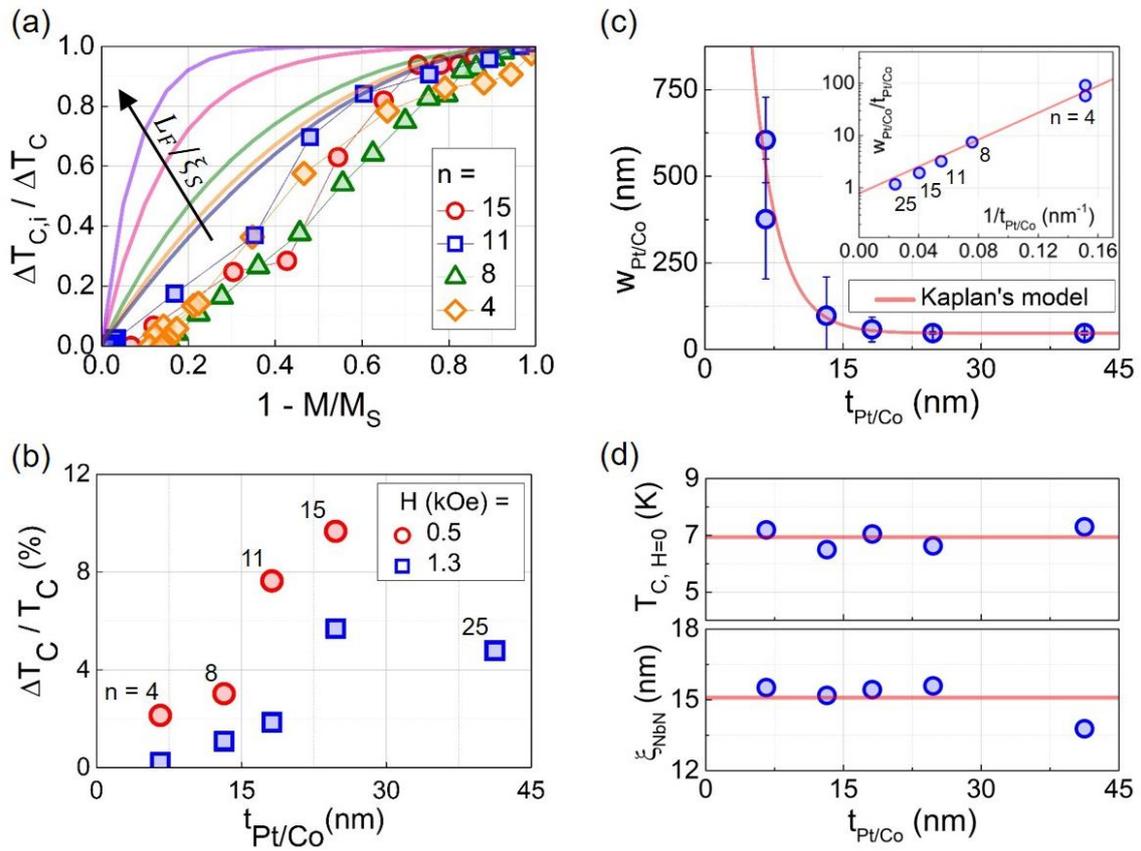

Fig. 4

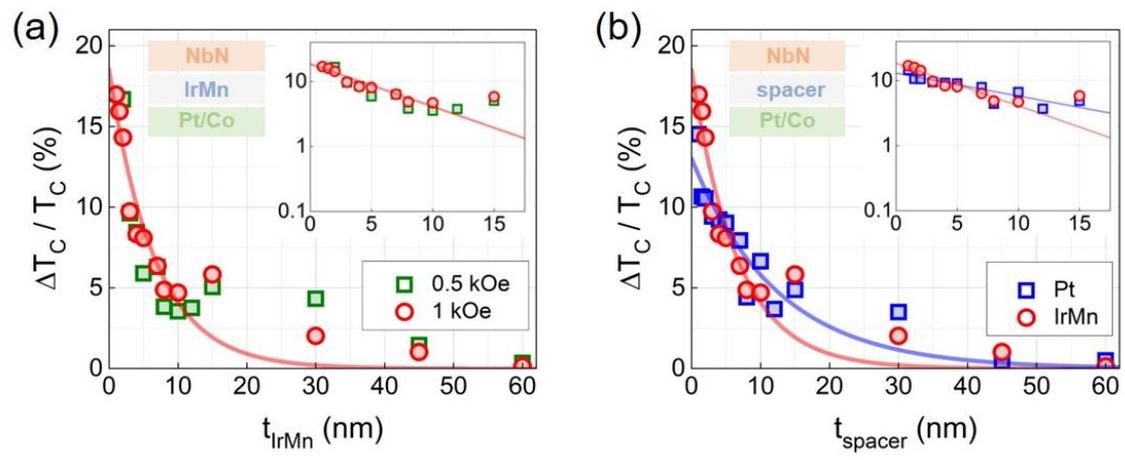

Fig. 5

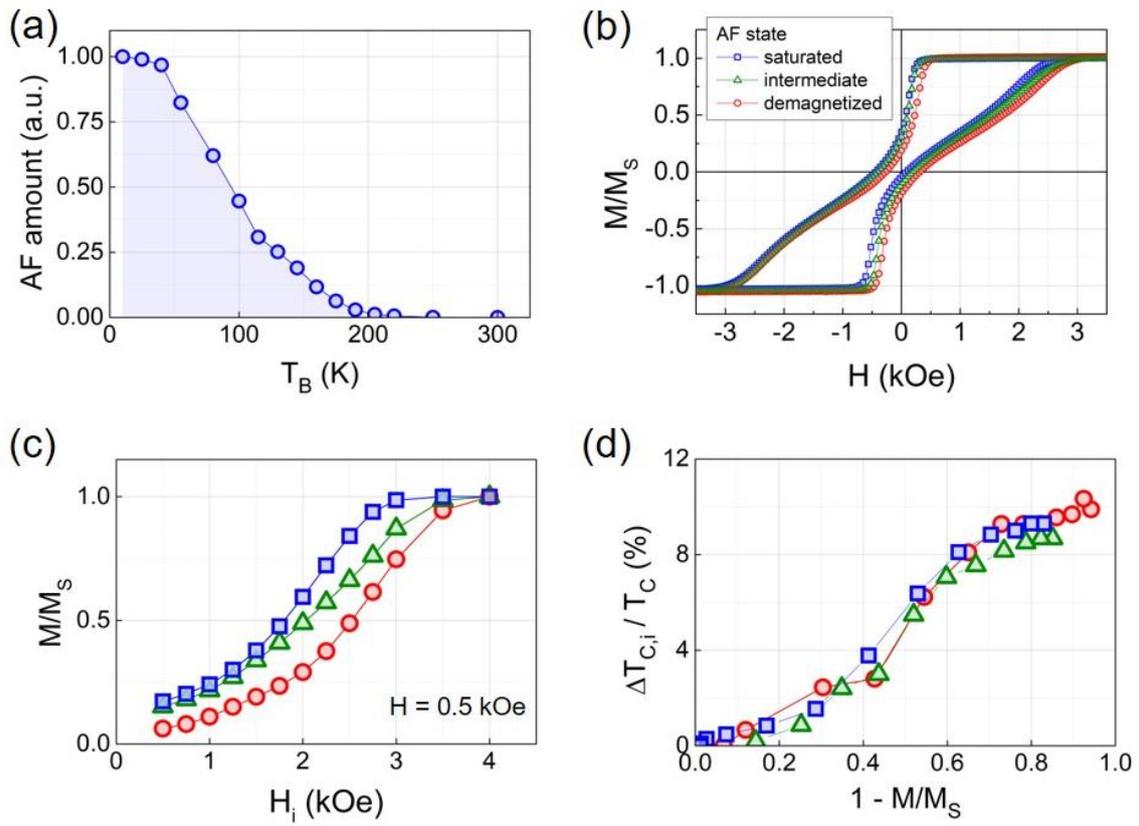

Fig. 6